\documentclass[twocolumn,times,trackchanges]{aastex63}
\usepackage[utf8]{inputenc}

\begin{document}
\title{A Simple Code for Rotational Broadening of Broad Wavelength Range High-Dispersion Spectra}
\author[0000-0002-9540-853X]{Adolfo Carvalho}
\affiliation{Department of Astronomy; California Institute of Technology; Pasadena, CA 91125, USA}
\author[0000-0002-8828-6386]{Christopher M. Johns-Krull}
\affiliation{Department of Physics and Astronomy; Rice University; Houston, TX 77005, USA}

\begin{abstract}
We present a simple method for rotationally broadening broad wavelength ranges of high-dispersion spectra. The broadening is rapid and scales linearly with the length of the spectrum array. For large wavelength ranges, the method is much faster than the popular convolution-based broadening. We provide the code implementation of this method in a publicly accessible repository.

\end{abstract}
\keywords{high-dispersion spectroscopy, open-source code, rotational broadening}

\section{Introduction}
As faster and more intricate techniques for fitting and modeling high-dispersion spectra develop, one step in the calculation that is often overlooked is the rotational broadening of templates or models. A popular technique to rapidly broaden a small region of a spectrum is to convolve the spectrum by the kernel prescribed in \citet{gray_2005}. 

Unfortunately this method typically relies on a wavelength grid that is sampled at constant $d\lambda$ and computes the Doppler shifts in wavelength space. This is only valid for a small region of the spectrum, $\Delta v \sim 3000$ km s$^{-1}$, where the wavelength dependence of the Doppler shift is small and can be ignored. 

In order to apply the method to larger wavelength ranges accurately, the spectrum must be separated into many smaller $\Delta v \sim 3000$ km s$^{-1}$ arrays which are each broadened then reassembled into a single broadened spectrum. Even a single echelle order on Keck/HIRES or other similar high-dispersion spectrographs must be separated into at least two segments for accurate convolutional broadening. 

In this note we present a simple code which directly integrates the stellar (or planetary) disk, allowing for accurate rotational broadening across a broad wavelength range. 

\section{Broadening by direct integration}
Rotational broadening by direct integration is relatively simple and the time taken for the broadening depends mainly linearly on the length of the vector being broadened. 

We compute the broadening by projecting the spherical stellar surface onto the two-dimensional sky. We first segment the disk into polar coordinates, $r$ and $\theta$, with $r \in [0,1]$. The radial grid size is given by $N_r$, with the corresponding spacing $dr = 1/N_r$, and the maximum number of $\theta$ steps in the outermost ring is given by $N_\theta$. At each radial point, the number of $\theta$ steps $n_\theta(r) = r N_\theta$, rounded to the nearest integer,  and $d\theta = (2\pi)/n_\theta$. At each $r, \ \theta$ point, we then define the projected area by
\begin{equation}
    dA(r, \theta) = \frac{\pi (r + dr/2)^2 - \pi (r - dr/2)^2}{n_\theta}. 
\end{equation}
We also incorporate the a linear limb-darkening law, defined by the parameter $\epsilon$, which ranges from 0 to 1, by rescaling the areas such that 
\begin{equation}
\hat{dA} = dA (r,\theta) \times (1-\epsilon + \epsilon \cos(\arcsin(r))).
\end{equation}
We then compute the projected velocity by $v(r,\theta) = v \ r \ sin(\theta)$. To incorporate differential rotation, we adopt a solar-like differential rotation law by scaling the projected velocity according to: 
\begin{equation}
    \hat{v} = v(r, \theta) \times \left[1 - \frac{\delta}{2} - \frac{\delta}{2} \cos(2 \arccos(r \cos(\theta))) \right]
\end{equation}

We can then integrate the disk by:
\begin{equation}
    \hat{s}(\lambda) = \int^1_0 \int^{2\pi}_{0} s(\lambda(\hat{v}(r, \theta))) \hat{dA}(r, \theta),
\end{equation}
where $s(\lambda(\hat{v}(r, \theta)))$ is the spectrum interpolated onto the wavelength scale that has been Doppler shifted to the new velocity $\hat{v}$. We then normalize by 
\begin{equation}
    \hat{A} = \int^1_0 \int^{2\pi}_{0} \hat{dA}(r, \theta),
\end{equation}
to account for any under or over estimation of the disk area. We provide the code implementation of this method in \footnote{\url{https://github.com/Adolfo1519/RotBroadInt}}.

The accuracy of the method is a function of the resolution used in the disk integration ($N_r$ and $N_\theta$) and the length of the spectrum being broadened. We provide in the code the default $N_r$, $N_\theta$ values for which the broadening is computed accurately but still rapidly in most cases. 

\begin{figure*}[!htb]
    \centering
    \includegraphics[width=0.61\linewidth]{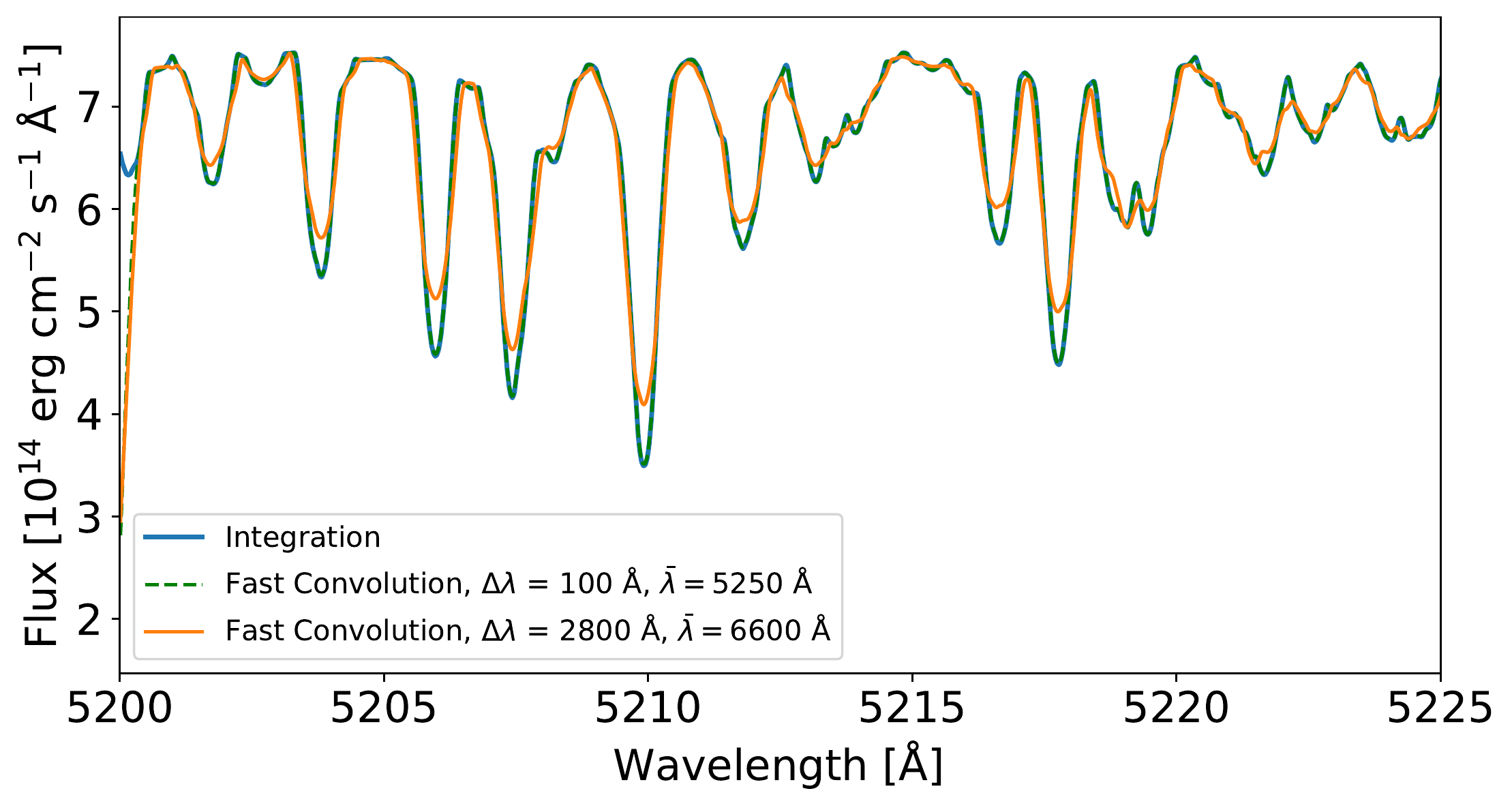}
    \includegraphics[width=0.35\linewidth]{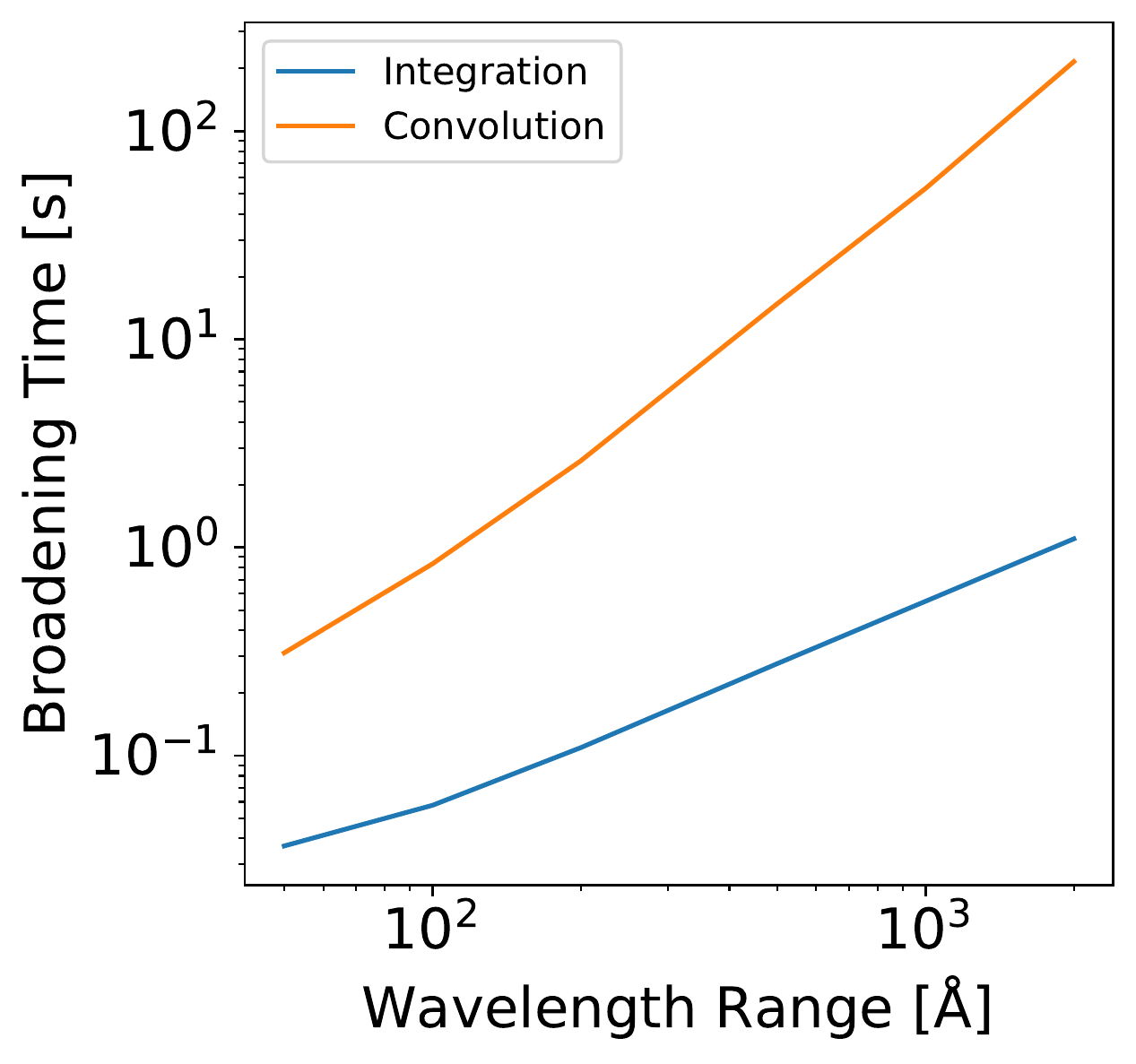}
    \caption{Results from rotationally broadening spectra using the two methods, direct integration and convolution. 
    \textbf{Left:} A 5500 K PHOENIX \citep{Husser_Phoenix_2013A&A} atmosphere model, broadened to 30 km s$^{-1}$ using the two methods and assuming a limb darkening coefficient of 0.6. We show a small piece of spectrum to better distinguish the individual line profiles. The wavelength ranges ($\Delta\lambda$) and mean wavelengths ($\bar{\lambda}$) of each spectrum are given in the legend. Notice the "fast" convolution matches the direct integration almost perfectly when the wavelength range is small and the region of interest is near the center of that range. In that case, the result is the same as using the "slow" method. However, for the larger wavelength range, where we are now far from the mean central wavelength, the broadening accuracy is much worse. The direct integration case is computed on the $\Delta \lambda = 2800 \ \mathrm{\AA}$ case. 
    \textbf{Right:} The times for the two broadening method as a function of the length of the array being broadened. Notice the rapid increase in time for the convolution method compared to integration. The spectra shown are sampled at $d\lambda \sim 0.01$, so $\Delta\lambda = 1000$ is a $10^5$ element array. The grid used for the integration has $N_r = 10$ and $N_\theta = 100$, the default values for our function.}
    \label{fig:rotbroadfig}
\end{figure*}

We show the result of broadening a $T_\mathrm{eff} =5500$ K PHOENIX \citep{Husser_Phoenix_2013A&A} atmosphere model spectrum using direct integration versus convolution in Figure \ref{fig:rotbroadfig}. We use the convolutional broadening code in the package $\mathtt{PyAstronomy}$ \citep{pya} for this example. For the sake of comparing the accuracy of the methods across broad wavelength ranges, we use the "fast" version of convolution, wherein the spectrum is not subdivided. We do this to demonstrate the inaccurate results it produces for regions of the spectrum far from the central wavelength adopted in the convolution.

For the smaller $\Delta \lambda$ shown in Figure \ref{fig:rotbroadfig}, the convolutional broadening produces an identical result to direct integration. In this case, the $\Delta \lambda$ is small enough to be valid for the kernel approximation and the "fast" convolution is essentially the same as the "slow" convolution. For the larger $\Delta \lambda$, the "fast" convolution is insufficient to accurately reproduce the rotational broadening.  

We also compare the computation times for direct integration compared with the more properly done "slow" convolution (subdividing the spectrum into smaller wavelength bins). For a fixed resolution, the computation time of integration scales approximately linearly with the length of the spectrum. This compares quite favorably with the exponential increase in computation time for convolution. 

For a high-dispersion spectrum sampled at $d\lambda < 0.1$, a wavelength range of $\Delta \lambda \sim 2000 \ \mathrm{\AA}$ has $\sim 20000$ points. The computation time of the convolutional broadening for an input this long is several minutes. While this may be acceptable for a single demonstration, it makes broadband fitting impossible. 

An additional benefit of our implementation of the direct integration is that it does not reply on having an evenly sampled wavelength grid. The convolution method requires that the spectrum being convolved has a constant $\Delta \lambda$ throughout. However, this is not possible if the spectra are sampled at constant $\Delta v$. If representing the data at a consistent resolution ($R = \lambda / \Delta \lambda$) across a broadband spectrum is important, the convolution method should not be used. 

\section{Conclusion}
We encourage the use of our public, open-source rotational broadening code for anyone doing broadband high-dispersion spectroscopy. The method is simple but powerful and effective.

\bibliography{references}{}

\begin{thebibliography}{}
\expandafter\ifx\csname natexlab\endcsname\relax\def\natexlab#1{#1}\fi
\providecommand{\url}[1]{\href{#1}{#1}}
\providecommand{\dodoi}[1]{doi:~\href{http://doi.org/#1}{\nolinkurl{#1}}}
\providecommand{\doeprint}[1]{\href{http://ascl.net/#1}{\nolinkurl{http://ascl.net/#1}}}
\providecommand{\doarXiv}[1]{\href{https://arxiv.org/abs/#1}{\nolinkurl{https://arxiv.org/abs/#1}}}

\bibitem[{{Czesla} {et~al.}(2019){Czesla}, {Schr{\"o}ter}, {Schneider},
  {Huber}, {Pfeifer}, {Andreasen}, \& {Zechmeister}}]{pya}
{Czesla}, S., {Schr{\"o}ter}, S., {Schneider}, C.~P., {et~al.} 2019, {PyA:
  Python astronomy-related packages}.
\newblock \doeprint{1906.010}

\bibitem[{Gray(2005)}]{gray_2005}
Gray, D.~F. 2005, The Observation and Analysis of Stellar Photospheres, 3rd
  edn. (Cambridge University Press), \dodoi{10.1017/CBO9781316036570}

\bibitem[{{Husser} {et~al.}(2013){Husser}, {Wende-von Berg}, {Dreizler},
  {Homeier}, {Reiners}, {Barman}, \& {Hauschildt}}]{Husser_Phoenix_2013A&A}
{Husser}, T.~O., {Wende-von Berg}, S., {Dreizler}, S., {et~al.} 2013, \aap,
  553, A6, \dodoi{10.1051/0004-6361/201219058}

\end{thebibliography}
\bibliographystyle{aasjournal}

\end{document}